\documentstyle[12pt,epsf]{article}

\begin{document}

\begin{titlepage}
\begin{flushright}
\vspace{-1.0cm}{\normalsize UTHEP-344\\
UTCCP-P-18\\
September 1996\\}
\end{flushright}

\vspace*{0.5cm}
\begin{centering}
{\Large \bf 
Scaling of Chiral Order Parameter\\
in Two-Flavor QCD\\
}

\vspace{1.8cm}

{\large 
Y.\ Iwasaki\rlap,$^{\rm a,b}$ K.\ Kanaya\rlap,$^{\rm a,b}$ 
S.\ Kaya\rlap,$^{\rm a}$ 
and T.\ Yoshi\'e$^{\rm a,b}$
}
\vspace{0.5cm}

{\it
$^{\rm a}$
Institute of Physics, University of Tsukuba,
Ibaraki 305, Japan\\
$^{\rm b}$
Center for Computational Physics, University of Tsukuba,\\
Ibaraki 305, Japan\\
}

\end{centering}

\vspace{1.5cm}\noindent
{
The finite temperature transition of QCD with two
degenerate light quarks is studied on the lattice
with a renormalization group improved gauge action
and the Wilson quark action.
We have made simulations
on an $8^3\times 4$ lattice near the chiral transition point.
It is shown that the chiral condensate which is the order parameter
of chiral symmetry satisfies remarkably a scaling relation
with the exponents of the three dimensional O(4) Heisenberg model.
This indicates that the chiral transition in two-flavor QCD is
of second order in the continuum limit.
}

\vfill \noindent

\end{titlepage}

Understanding the nature of the finite temperature transition 
in QCD with two degenerate light quarks ($N_F=2$) is 
an important step toward the clarification of the transition 
in the real world. 
It is plausible from an argument based on an effective $\sigma$ model
\cite{Wilczek}
that if the chiral transition (transition in the chiral limit)
is of second order, QCD with two flavors
belongs to the same universality class as the three 
dimensional O(4) Heisenberg model.
This universality provides us with several useful scaling relations
that can be confronted with numerical results.
These scaling relations were 
first tested on the lattice for staggered quarks 
\cite{KarschLaermann}, 
where some evidences consistent with the O(4) scaling were reported. 

In this paper, we study scaling behavior of
the QCD transition 
with two degenerate light quarks on the lattice 
using the Wilson formalism of fermions,
and compare the result with a scaling relation conjectured.
The Wilson formalism of fermions 
is the only known lattice formalism which possesses a local action 
for any number of flavors.  Therefore, it is important to investigate 
the scaling behavior with Wilson quarks and compare the results 
with those 
for staggered quarks for which the action is non-local for two flavors.

We use a renormalization group improved gauge action \cite{IM} 
which is expected to be closer to the renormalized trajectory 
compared with the standard one-plaquette gauge action, and therefore
is expected to show scaling behavior on a coarse lattice.

The continuum limit does not depend on the choice of the action
if the action belongs to the same universality class.
However, the way of the approach to the continuum limit does depend 
on the choice of the action. 
When one uses the standard one-plaquette gauge action and the Wilson
quark action for a study of QCD with two flavors at finite 
temperatures,
one encounters the existence of severe 
lattice artifacts on lattices with the lattice size in the euclidean 
time direction $N_t=4$ and 6
\cite{MILC46,IwasakiKanaya,ourStandard96}.
Note the temperature on the lattice is given by $T=1/(a N_t)$ with
$a$ being the lattice spacing.
A way out of these lattice artifacts with the standard action 
is to increase $N_t$ 
in order to further approach toward the continuum limit. 
However, our previous study on an $18^3\times24$ lattice 
suggests that $N_t > 18$ is required to have the
chiral transition in the scaling region 
\cite{ourLatOld}.
It is difficult to perform simulations on such a large lattice 
even with a powerful computer of today.
Therefore, we instead adopt 
a renormalization group improved gauge action
and the Wilson quark action.

Before going into the discussion of the present work,
let us first describe unexpected phenomena observed 
by the MILC collaboration with the standard gauge action.
The transition is smooth both for heavy and light quarks,
while the transition is very sharp at $N_t=4$ 
(even a first order transition at $N_t=6$)
in the range of intermediate quark masses ($\beta \simeq 5.0$)
\cite{MILC46}. 
This is completely different from what is supposed to be realized
in the continuum limit:
as the quark mass increases from the chiral limit,
the transition becomes weaker 
and it becomes strong again only when the quark mass is heavy enough 
to recover the first order transition of the SU(3) gauge theory.
They also observed a cusp in the behavior of $m_\pi^2$ in terms of 
$1/K$ with $\beta$ fixed ($m_\pi$ is the screening pion mass 
and $K$ is the hopping parameter).
Furthermore,
at $\beta$
{\raise0.3ex\hbox{$<$\kern-0.75em\raise-1.1ex\hbox{$\sim$}}}
5.3,
the value of the quark mass $m_q$, defined through
an axial-vector Ward identity \cite{Bochicchio,ourMq},
depends on the phase of the system
and shows unexpected $1/K$ dependence with $\beta$ fixed
in the high temperature phase \cite{MILC46,YoshieLat9495},
which is in sharp contrast with the case of $\beta$
{\raise0.3ex\hbox{$>$\kern-0.75em\raise-1.1ex\hbox{$\sim$}}}
5.5 where it does not depend on the phase \cite{ourMqQ}.
In the following, we show that these unexpected phenomena are 
removed by improving the lattice action.

Now let us discuss the present work.
The renormalization group improved gauge action $S_q^{\rm IM}$ 
we use is given by
\begin{equation}
S_g^{\rm IM} = {1 \over g^2}\, \{c_0 \sum (1\times 1 {\rm \ loop}) 
               + c_1 \sum (1\times 2 {\rm \ loop})\}
\label{eq:Sg}
\end{equation}
with $c_1=-0.331$ and $c_0=1-8c_1$ \cite{IM}.
Here the loops in the sums are defined by the trace of ordered 
product of 
link variables, and each oriented loop appears once in the sum. 

We mainly perform simulations for $N_t=4$ with
spatial lattice size $8^3$
at $\beta \equiv 6/g^2 = 1.0$ --- 4.0
near the finite temperature transition point.
We use an anti-periodic boundary condition in the $t$ direction
and periodic boundary conditions otherwise.
The configurations are updated using the hybrid Monte Carlo 
method 
with the molecular dynamics time step size $\delta\tau=0.01$ 
(except for a simulation point $\beta=1.5$, $K=0.206$ on the 
$N_t=4$ lattice where $\delta\tau=0.002$). 
We perform simulations of about 100 --- 1300 trajectories after 
thermalization for each set of parameters, $\beta$ and $K$.
Errors are estimated by the jackknife method. 
A preliminary report is given in Ref.\cite{YoshieLat9495}. 

\begin{figure}[tb]
\epsfxsize=13.5cm \epsfbox{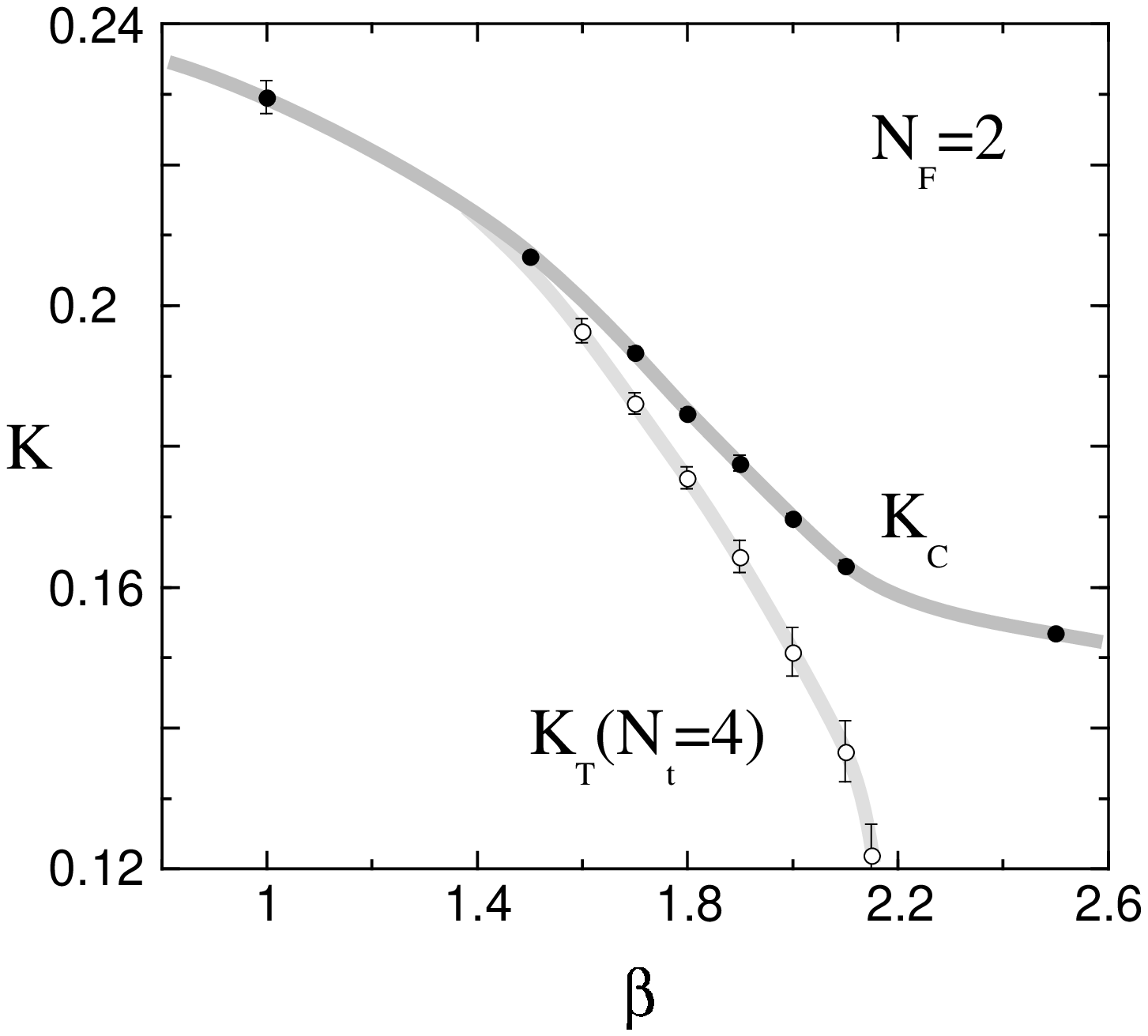}
\caption{
Phase diagram for $N_F=2$ QCD with the RG improved gauge action
Eq.(\protect\ref{eq:Sg}) and the Wilson quark action.
Lines are to guide the eyes.
}
\label{fig:iKcKt_F2}
\end{figure}

\begin{figure}[tb]
\begin{center}\leavevmode
\epsfxsize=10cm \epsfbox{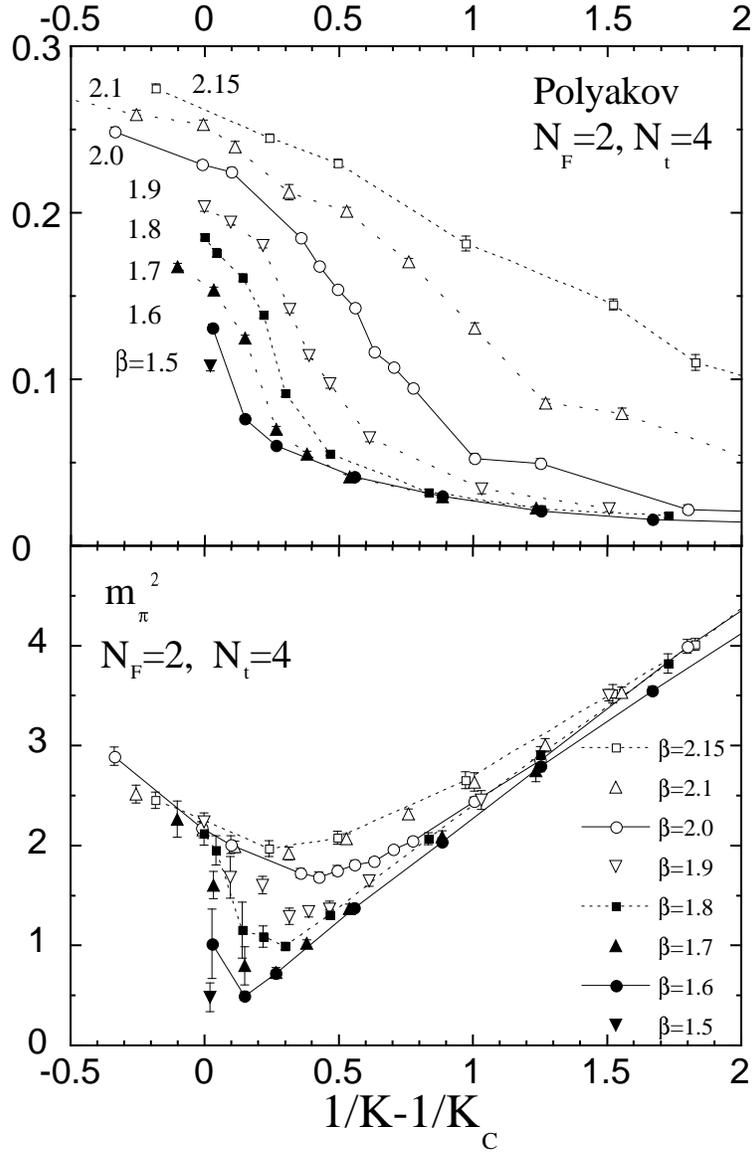}
\end{center}
\caption{
The Polyakov loop and the pion screening mass. 
}
\label{fig:iF2T4PMpi}
\end{figure}

Our results for the phase diagram are shown in 
Fig.~\ref{fig:iKcKt_F2}. 
The chiral line $K_c$ \cite{ourStandard96} is defined by 
the vanishing point of the screening pion mass, $m_\pi = 0$, 
on an $8^4$ lattice with periodic boundary conditions
(doubled in the $t$ direction for the spectrum calculation).
The finite temperature transition/crossover line $K_t$ for 
$N_t=4$ is determined on the $8^3 \times 4$ lattice
from the condition that $P=0.10(2)$ where
$P$ is the Polyakov line expectation value. 
This condition corresponds to the 
criterion that the crossover point is identified as
the peak position of the susceptibility of the Polyakov loop, 
although our limited statistics sometimes makes the peak of 
the susceptibility not so clear. 
We also note that this is consistent with the criterion
that the crossover points are identified from
rapid changes of physical quantities 
(see Fig.~\ref{fig:iF2T4PMpi}). 
We identify the crossing point of the $K_c$ and $K_t$ lines as
the point of the finite temperature
chiral transition point, 
$\beta_{ct}$ (cf.\ Ref.\cite{ourStandard96}).

For the case of the standard gauge action, 
we pointed out that the simulation points where the strong 
transitions mentioned earlier 
are observed on the $K_t$ line \cite{MILC46} are just
those where the $K_t$ line 
approaches toward the $K_c$ line after initially deviating from it
due to the cross-over phenomenon between weak and 
strong coupling regions of QCD,  
and that therefore it seems plausible that the strong transition is 
a result of lattice artifacts 
caused by this unusual relation of the $K_t$ and $K_c$ lines
\cite{IwasakiKanaya,ourStandard96,YoshieLat9495}. 

Unlike for the case of the standard gauge action,
the distance between the $K_t$ and $K_c$ lines 
shown in Fig.~\ref{fig:iKcKt_F2} grows monotonically 
when we increase $\beta$ from the chiral transition point 
$\beta_{ct} \sim 1.4$. 
In accord with this, 
the transition becomes monotonically weaker with $\beta$
as shown in Fig.~\ref{fig:iF2T4PMpi}a for the Polyakov loop.
This is in sharp contrast with the case of the standard gauge action.
Note that the change of the Polyakov loop 
at the finite temperature transition/crossover
is very smooth for a wide range of $\beta$.
The change of $m_\pi^2$ is also smooth for larger values of $\beta$
and it becomes sharper as $\beta$ decreases 
(cf.\ Fig.~\ref{fig:iF2T4PMpi}b).
The straight line envelop of $m_\pi^2$ in the high temperature
phase ($N_t=4$) agrees with $m_\pi^2$ in the low temperature phase 
($N_t=8$),
and corresponds to the PCAC relation $m_\pi^2 \propto m_q$.
The smoothness of physical observables strongly suggests that 
the transition is a crossover at $\beta > \beta_{ct}$. 

\begin{figure}[tb]
\epsfxsize=13.5cm \epsfbox{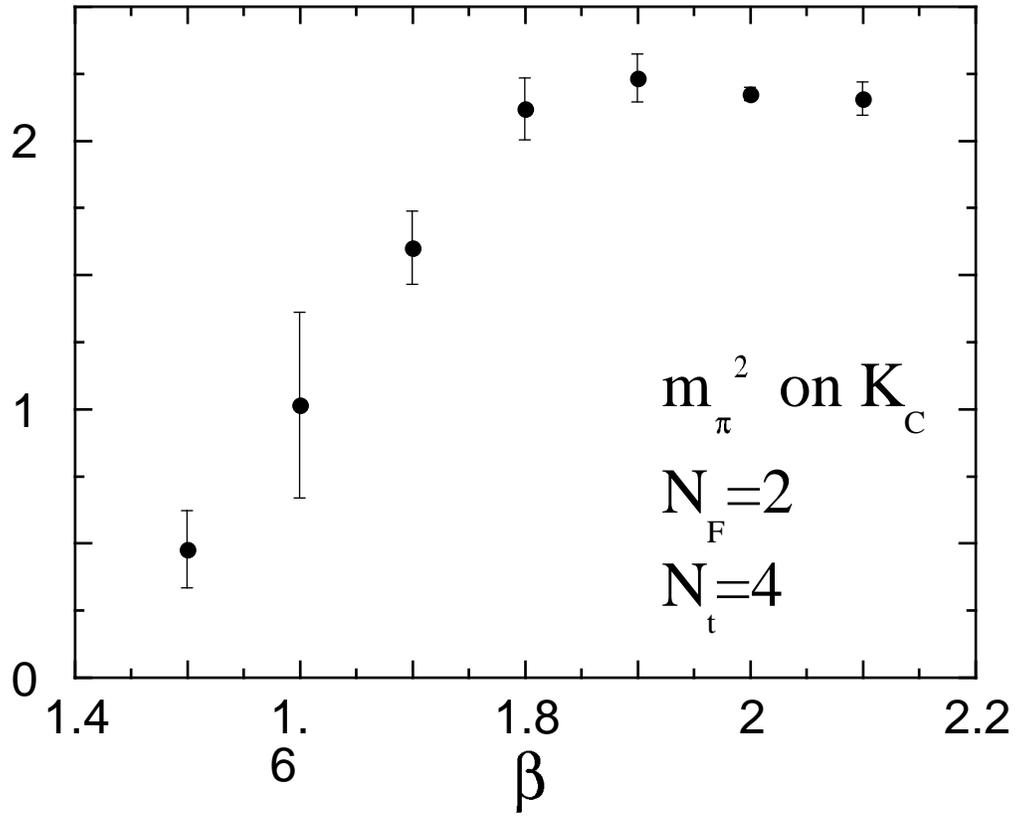}
\caption{
$m_{\pi}^2$ on $K_c$. 
}
\label{fig:iF2KcPi2}
\end{figure}

\begin{figure}[tb]
\epsfxsize=13.5cm \epsfbox{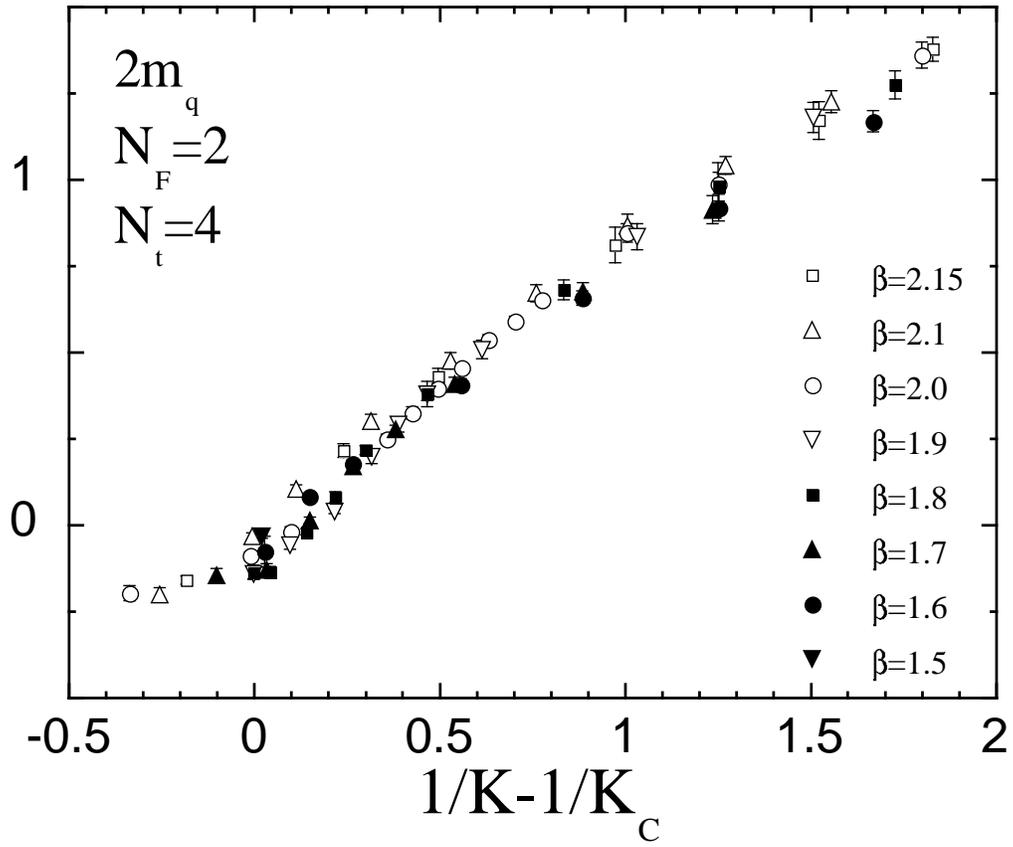}
\caption{
The same as Fig.~\protect\ref{fig:iF2T4PMpi} 
for twice the quark mass.
}
\label{fig:iF2T4Mq}
\end{figure}

It should be noted that 
the value of $m_\pi^2$ on the $K_c$ line monotonically
decreases to zero as $\beta \rightarrow \beta_{ct} + 0$
(cf.\ Fig.~\ref{fig:iF2KcPi2}), 
suggesting that the chiral transition is continuous. 
We also note that $m_\pi^2$ on the $K_t$ line shows a similar 
monotonic decrease \cite{YoshieLat9495}. 
It might be emphasized that the value of $m_q$ in the 
high temperature phase agree well with that in the low temperature
phase and does not show the strange behavior
mentioned before for the case of the 
standard action (see Fig.~\ref{fig:iF2T4Mq}).

These nice properties 
which are in accordance with naive expectations 
encourage us to begin a scaling study with Wilson quarks. 

From the universality argument 
we expect that magnetization $M$ near the second order phase
transition point can be described by a single scaling function: 
\begin{equation}
M / h^{1/\delta} = f(t/h^{1/(\beta\delta)})
\label{eq:universality}
\end{equation}
where $h$ is the external magnetic field 
and $t=[T-T_c(h\!=\!0)]/T_c(h\!=\!0)$ is the reduced temperature. 
For three dimensional O(4) models, the critical exponents 
in (\ref{eq:universality}) are given by 
$1/(\beta\delta) = 0.537(7)$ and $1/\delta = 0.2061(9)$ 
\cite{KanayaKaya}.
In QCD, $h$ corresponds to the quark mass and 
$M$ corresponds to the chiral condensate. 
DeTar tested this scaling for the case of 
two flavor staggered quarks \cite{DeTarLat94} 
and reported that data are consistent with O(4) and O(2) scaling. 

\begin{figure}[tb]
\epsfxsize=13.5cm \epsfbox{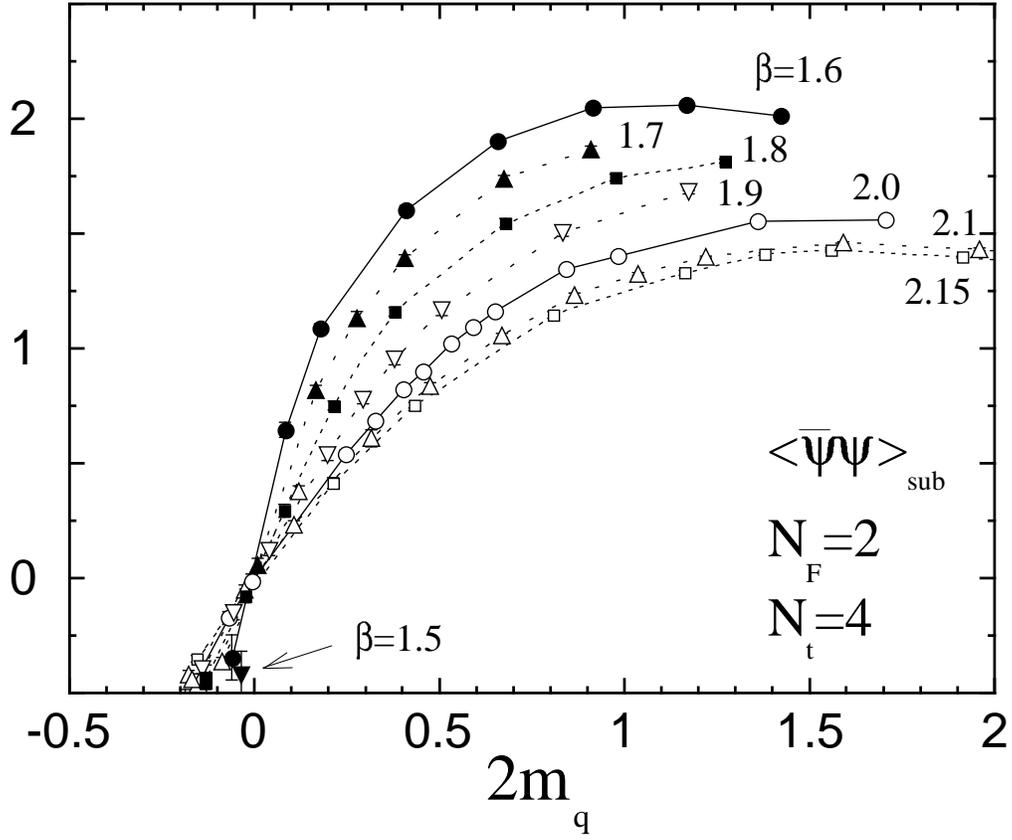}
\caption{
Subtracted chiral condensate 
$\langle \bar{\Psi} \Psi \rangle_{\rm sub}$
as a function of $2m_qa$. 
}
\label{fig:iF2T4PBPsub}
\end{figure}

For Wilson quarks, 
the naive definition of $\langle \bar{\Psi} \Psi \rangle$
for the chiral condensate 
is not adequate because the chiral symmetry is explicitly 
broken due to the presence of the Wilson term. 
A proper subtraction and a renormalization are required to obtain 
the correct continuum limit. 
A properly subtracted 
$\langle \bar{\Psi} \Psi \rangle$ can be defined via an axial 
Ward identity \cite{Bochicchio}:
\begin{equation}
\langle \bar{\Psi} \Psi \rangle_{\rm sub} 
= 2 m_q a Z \sum_x \langle \pi(x) \pi(0) \rangle
\label{eq:PBPsub}
\end{equation}
where $Z$ is the renormalization coefficient. 
This definition is 
consistent with the identification of the magnetization 
in Ref.\cite{Wilczek}. 
$\langle \bar{\Psi} \Psi \rangle_{\rm sub}$ 
was shown to have a non-vanishing value in the chiral limit 
in the confining phase of quenched QCD \cite{PBPsubQCD}. 
For our purpose, it is enough to use the tree value: 
$Z=(2K)^2$. 
Our results of 
$\langle \bar{\Psi} \Psi \rangle_{\rm sub}$ for $N_t=4$ 
are shown in Fig.~\ref{fig:iF2T4PBPsub}. 

\begin{figure}[tb]
\begin{center}\leavevmode
\epsfxsize=10cm \epsfbox{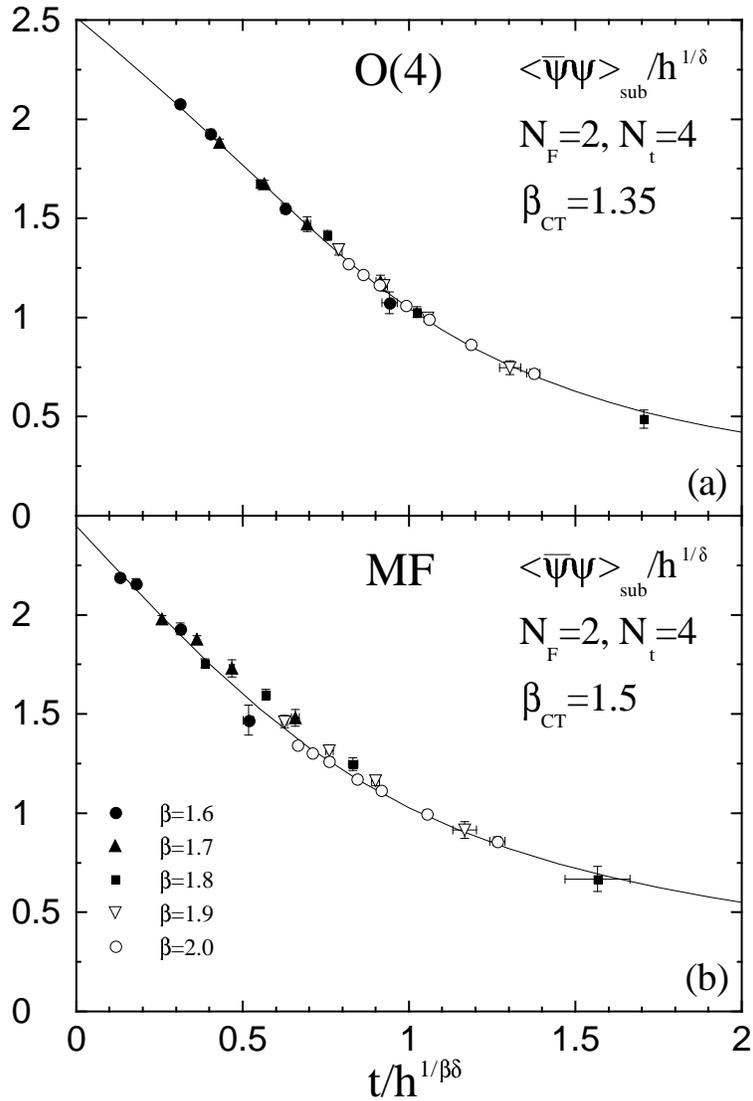}
\end{center}
\caption{
Best fits to the scaling function 
with (a) O(4) and (b) MF exponents (see text). 
The plot contains all the data shown in 
Fig.~\protect\ref{fig:iF2T4PBPsub}
within the range $0 < 2m_q a < 0.8$ and $\beta \leq 2.0$. 
Solid curves are scaling functions obtained in an O(4) 
spin model \protect\cite{Toussaint} 
and by a MF calculation, respectively. 
}
\label{fig:PBPscaling}
\end{figure}

If the two flavor QCD belongs to the same universality class 
as three dimensional O(4) spin models, the chiral condensate 
should satisfy the scaling relation (\ref{eq:universality}) 
with the identification 
$M=\langle \bar{\Psi} \Psi \rangle_{\rm sub}$, $h=2m_q a$, 
and $t=\beta-\beta_{ct}$. 
We further expect that the scaling function $f(x)$ itself is 
a universal function \cite{Toussaint} 
because $f(x)$ is determined by the universal singular structure of 
the free energy around the UV fixed point.
We make a fit to the scaling function
recently obtained for an O(4) model \cite{Toussaint},
by adjusting $\beta_{ct}$ and the scales for $t$ and $h$, 
with the exponents fixed to the O(4) values.
Fig.~\ref{fig:PBPscaling}a shows our result
with $\chi^2/df = 0.61$.
The scaling ansatz works remarkably well with the O(4) exponents.
The resulting $\beta_{ct} = 1.35(1)$ is slightly smaller than 
the value $\simeq 1.4$ 
obtained by linear extrapolations of the $K_t$ line
in the $(\beta,K)$ space (cf.\ Fig.~\ref{fig:iKcKt_F2}),
the $m_\pi^2$ on $K_c$ (cf.\ Fig.~\ref{fig:iF2KcPi2}), 
and also the $m_\pi^2$ on $K_t$ \cite{YoshieLat9495}. 
However, the O(4) universality predicts \cite{Wilczek} that 
$\beta_c(m_q) - \beta_{ct} \propto m_q^{1/(\beta\delta)}$,
and $m_\pi^2 \sim (\beta_c-\beta_{ct})^\gamma$ 
on the $K_c$ and $K_t$ lines
with $\gamma \simeq 1.4$, which implies that these lines should bend 
slightly near the chiral transition point 
to give a smaller $\beta_{ct}$. 
Therefore, we conclude that $\beta_{ct} = 1.35(1)$ is consistent 
with the data. 

Finally, we test the meanfield (MF) exponents 
[$1/(\beta\delta) = 2/3$, $1/\delta = 1/3$]
suggested in Ref.\cite{KocicKogut}. 
We find that it is more difficult to arrange a wide range of data
on a universal curve by adjusting $\beta_{ct}$.
A fit with the minimum $\chi^2/df = 2.5$ with the MF 
scaling function is obtained with $\beta_{ct}=1.54(1)$. 
However, the data at $\beta=1.6$ are completely off the fit.
Furthermore, this value of $\beta_{ct}$ is in conflict with the fact
that the data of $m_\pi^2$ on the $K_c$ line indicates 
$\beta_{ct} \le 1.5$ (cf.\ Fig.~\ref{fig:iF2KcPi2}). 
Restricting $\beta_{ct} \le 1.5$, we obtain the fit shown in 
Fig.~\ref{fig:PBPscaling}b with $\chi^2/df = 3.3$. 
In contrast with the case of the O(4) exponents, 
the MF scaling function can not well reproduce the
$t/h^{1/(\beta\delta)}$ dependence of the data 
at fixed $\beta$, especially at small $\beta$'s. 
As a result, the whole data are more scattered 
than the case of the O(4) exponents. 
When we decrease $\beta_{ct}$ less than 1.4, 
which seems natural from other data 
(cf.\ Figs.~\ref{fig:iKcKt_F2}, \ref{fig:iF2KcPi2}),
$\langle \bar{\Psi} \Psi \rangle_{\rm sub} / h^{1/\delta}$
does not fall on an approximate universal curve. 
Therefore, we conclude that the data do not favor the MF scaling.

The success of this scaling test with the O(4) exponents
strongly suggests 
that the chiral transition is of second order 
in the continuum limit. 
It also indicates that, with the improved gauge action, 
the chiral violation due to the Wilson term 
is sufficiently weaker 
than that introduced by the non-vanishing $m_q$ 
at least for $m_q$'s at $\beta$'s studied here. 
To strengthen the conclusion, 
a direct extraction of each critical exponent 
would be desirable.

We are grateful to Doug Toussaint for very useful discussions and 
sending us the data of the scaling function for an O(4) spin model.
The simulations are performed with Fujitsu VPP500/30 at
the University of Tsukuba.
This work is in part supported by the Grant-in-Aid
of Ministry of Education, Science and Culture
(Nos.07NP0401, 07640375 and 07640376).

\clearpage

\end{document}